\documentclass[preprint,twocolumn,pra,10pt,aps,amsmath,amssymb,nofootinbib,superscriptaddress,showkeys,showpacs]{revtex4-1}
\usepackage{epsfig} 
\usepackage{amsmath} 
\usepackage{graphicx}
\usepackage{color}
\usepackage{float}
\newcommand{\ba}{\begin{array}}
\newcommand{\ea}{\end{array}}
\newcommand{\bea}{\begin{eqnarray}}
\newcommand{\eea}{\end{eqnarray}}

\def\bra#1{\left\langle #1\right|}
\def\ket#1{\left| #1\right\rangle}

\begin{document} 
\preprint{UdeM-GPP-TH-14-237}

\title{Quantum correlations in $B$ and $K$ meson systems}

\author{Subhashish Banerjee}
\email{subhashish@iitj.ac.in}
\affiliation{Indian Institute of Technology Jodhpur, Jodhpur 342011, India}

\author{Ashutosh Kumar Alok}
\email{akalok@iitj.ac.in}
\affiliation{Indian Institute of Technology Jodhpur, Jodhpur 342011, India}

\author{Richard MacKenzie}
\email{richard.mackenzie@umontreal.ca}
\affiliation{Universit\'e de Montr\'eal,
Pavillon Roger-Gaudry, D\'epartement de physique,
C.P. 6128, Succursale Centre-ville,
Montr\'eal, Qc    H3C 3J7
Canada}

\date{\today}

\begin{abstract}
The interplay between the various measures of quantum correlations are well known in stable optical and electronic systems. Here, for the first time, we study such foundational issues in unstable quantum systems. Specifically we study meson-antimeson systems, which are produced copiously in meson factories. We use the semigroup formalism to compute the time evolution of several measures of quantum correlations for three meson systems ($K\bar K$, $B_d \bar{B}_d$ and $B_s \bar{B}_s$), circumventing difficulties which arise using other methods due to the instability of these particles. We then compare these measures to one another and find that the relations between them can be nontrivially different from those of their stable counterparts such as neutrinos.
\end{abstract}


\maketitle 

\section{Introduction}

Quantum correlations in measurements performed on multipartite systems provide a fertile testing ground for foundational aspects of quantum physics. They are also of central importance to potential applications such as quantum communication, computation and cryptography. They have been studied and applied in a variety of physical contexts such as quantum optics and condensed matter systems (superconductors, spin systems, {\em etc.}). More recently, attention has also been directed towards subatomic physics 
\cite{Bramon:1998nz,Bramon:2001tb,Bramon:2002yg,Bertlmann:2002wv,Bramon:2004zp,Hiesmayr:2007he,Bertlmann:2004cr,active,Blasone:2007vw,Blasone:2007wp,AmelinoCamelia:2010me,Hiesmayr:2011na,Blasone:2014jea,asu}, 
inspired by the technical advances in high energy physics experiments, in particular the meson factories, reactor and accelerator neutrino experiments.

The foundations of quantum mechanics are usually studied in optical or electronic systems, where  the interplay between the various measures of quantum correlations is well know\cite{louisell,bp,nc,path,hor,sb1}. In these contexts, the detection efficiency is much lower than that of the corresponding detectors in high energy physics experiments, such as the meson factories. It is therefore interesting and, as we will see, fruitful to test the foundations of quantum mechanics in unstable massive systems, such as the correlated $B\bar B$ and $K\bar K$ systems,  for which the interplay between the various measures of quantum correlations have not yet been studied. 

Of course, testing quantum correlations requires a bipartite (or multipartite) system. $B$ factories, electron-positron colliders tailor-made to study the production and decay of $B$ mesons, and $\phi$ factories, which perform the same function for $K$ mesons, provide an ideal testing ground. In the case of $B$ factories, the collider energy is tuned to the $\Upsilon$ resonance, so the first stage of the process is $e^+e^-\to\Upsilon$. The $\Upsilon$ then decays into $b\bar b$; these form a $B_q \bar{B}_q$ ($q=s,d$) pair through hadronization. All this happens essentially instantaneously. The $B$ mesons then fly apart and decay on a much longer time scale. An important feature of these systems for the study of correlations is the oscillations of the bottom and strangeness flavors $b\leftrightarrow s$, giving rise to $B\bar B$ oscillations, which have been crucial to the study of $CP$ violation in these systems.  Here we study a number of well-established measures of quantum correlations in the $B$ and $K$ meson systems. Such a study is complicated by the inherently unstable nature of these particles, as a result of which the standard Weisskopf-Wigner approach to time evolution \cite{Weisskopf:1930au,Weisskopf:1930ps} results in ambiguities due to probability loss caused by the decrease of the trace of the density matrix.  

The treatment of unstable systems has a long and distinguished history \cite{Weisskopf:1930au,Weisskopf:1930ps,
peierls-1,peierls-2,khalfin,prig,kabir,sud,Ellis:1992dz,Huet:1994kr,benatti,benatti2}, and continues to be of interest 
today \cite{Urbanowski:2014tra}. The requirement of causality prohibits a physical state with complex energy \cite{peierls-1}, 
while the study of decay of unstable states is facilitated by the use of the density matrix formalism. This view is supported 
by work on decaying systems using dynamical semigroups \cite{rajgopal}, where deep theorems from analysis, such as the 
Sz.-Nagy dilation theorem \cite{nagy}, were used to show that a unitary evolution of an unstable system along with its decay 
products is not feasible. Thus a decaying system is intrinsically an open system, even without explicitly invoking 
an external environment, and as a result it can have surprises not seen in its stable counterpart. This work \cite{rajgopal} also motivates the 
development of a feasible theory of such systems using a semigroup approach, which in turn is related to a superselection rule 
with respect to time reversal symmetry. The Langer-Sz.-Nagy-Foias theorem \cite{nagy} theorem is used to provide the splitting 
of a semigroup generator into unitary and non-unitary parts.

In this work we make use of the probability-preserving formalism of decaying systems \cite{rajgopal,caban} to study various 
measures of quantum correlations in $B$ and $K$ mesons. 
We find that the quantum correlations for these unstable systems can be {\it nontrivially different} from their stable 
counterparts.  
Also, it is not obvious how to perform experiments testing nonlocality, as quantified by Bell's inequality, for $B$ mesons due
to the absence of {\em active} measurements \cite{active,Bertlmann:2004cr}. Despite this hurdle,  
the framework of open quantum systems \cite{bp,path, alicki}, adapted to the probability-preserving 
formalism of decaying systems, enables us to make quantitative statements about Bell inequality violations for correlated 
neutral $K$ and $B$ mesons. The theory of open quantum systems asserts that any real system interacts with its environment 
(here, fluctuations of the quantum mechanical vacuum), resulting in loss of quantum coherence and the
transformation from pure to mixed states \cite{Hawking:1982dj}. Thus this work elucidates fundamental aspects of 
quantum mechanics of correlated neutral meson systems, and more generally of unstable quantum systems.

The notion of quantum entanglement was implicit in the seminal 1935 paper of Einstein, Podolsky and Rosen \cite{epr}.
It was Schr{\"o}dinger 
who later that year first coined the term ``entanglement" in a series of papers wherein he also introduced his eponymous cat 
thought experiment \cite{schr}. The subject was further developed with the conception of experimental tests of 
quantum mechanics {\em vs.} hidden-variable theories, in particular Bell's inequality \cite{bell} and refinements resulting 
in the Bell-CHSH (Clauser-Horn-Shimony-Holt) inequalities \cite{chsh}.
Until recently, entanglement was considered synonymous with quantum correlations in that it was thought that one necessarily 
implied the other. With the advent of quantum discord \cite{ollivier, henderson,luo,girolami1}, which quantifies the 
difference between the quantum generalizations of two classically equivalent formulations of mutual information, it became 
clear that quantum correlations are broader than entanglement. In general, it is very difficult to obtain an analytical 
formula for quantum discord because it involves an optimization over local measurements, requiring numerical
methods \cite{huang}. To overcome this difficulty, another measure of quantum correlation called geometric discord was 
introduced in \cite{dakic}. This quantifies the amount of non-classical correlation of an arbitrary quantum composite system 
in terms of its minimal distance from the set of classical states. Along with these measures, another important facet of 
quantum correlations is its operational aspect. Teleportation fidelity occupies an important place here and was developed to 
provide an operational meaning to entanglement \cite{hor,ben}.

The plan of this work is as follows. In the next section we set up the two-meson system, treating each meson as a state in a Hilbert space consisting of one-particle and zero-particle sectors. The relevant tool for computing quantum correlations is the density matrix projected down to the two-particle sector. In the following section various correlation measures are computed. We then summarize our results and highlight the main surprising result, which is that the relations between the correlation measures we consider can be different from the corresponding relations for stable systems.

\section{$M\bar M$ as an open quantum system}

For the $B$ system, imagine the decay $\Upsilon\to b\bar b$ followed by hadronization into a $B\bar B$ pair. In the $\Upsilon$ rest frame, the mesons fly off in opposite directions (left and right, say); since the $\Upsilon$ is a spin-1 particle, they are in an antisymmetric spatial state. The same considerations apply to the $K$ system, with the $\Upsilon$ replaced by a $\phi$ meson.

The flavor-space wave function of the correlated $M\bar{M}$ meson systems ($M=K,\, B_d, \,B_s$) at the initial time $t=0$ is
\begin{equation}
 \ket{\psi (0)} = \frac{1}{\sqrt{2}} \left[\ket{M \bar{M}} -\ket{\bar{M} M} \right], \label{flav}
\end{equation}
where the first (second) particle in each ket is the one flying off in the left (right) direction and $\ket{M}$ and $\ket{\bar{M}}$ are flavor eigenstates. As can be seen from \eqref{flav}, the initial state of the neutral meson system is a singlet (maximally entangled) state. The usual analysis of such systems is done using a trace-decreasing density matrix description of the state (see for example \cite{benatti}). However, such an approach may not be very useful for calculating quantum correlations, the subject of interest in this work. This is because the usual methods for computing quantum correlations require a trace-preserving, completely positive description of the system. The semigroup formalism enables the calculation of a trace-preserving density matrix. We also incorporate the effects of decoherence in our calculations, providing a uniform formalism for studying correlations in neutral meson systems.

The Hilbert space of a system of two correlated neutral mesons, as in \eqref{flav}, is  
\begin{eqnarray}
\mathcal{H}=(\mathcal{H}_{L}\oplus\mathcal{H}_0) \otimes (\mathcal{H}_{R}\oplus\mathcal{H}_0)\,,
\label{hilbert}
\end{eqnarray}
where $\mathcal{H}_{{L,R}}$ are the Hilbert spaces of the left-moving and right-moving decay products, each of which can be either a meson or an anti-meson, and $\mathcal{H}_0$ is that of the zero-particle (vacuum) state. Thus the total Hilbert space can be seen to be the tensor sum of a two-particle space, two one-particle spaces, and one  zero-particle state. The initial density matrix of the full system is 
\begin{equation}
\label{rhoh}
\rho_{\mathcal{H}}(0) = \ket{\psi (0)}\bra{\psi(0)}.
\end{equation}

The system, initially in the two-particle subspace, evolves in time into the full Hilbert space, eventually (after the decay of both particles) finding itself in the vacuum state. As can be appreciated from basic notions of quantum correlations such as entanglement, one needs to have two parties to correlate. For this we need to project from the full Hilbert space \eqref{hilbert} down to the two-particle sector $\mathcal{H}_{L}\otimes \mathcal{H}_{R}$, resulting in the following density matrix for the correlated neutral meson system:
\begin{equation}
 \rho (t) = \frac{1}{4}
\left(\begin{array}{cccc} 
a_- & 0 & 0 & -a_-  \\ 0 & a_+ & -a_+  & 0 \\ 
0 &-a_+   & a_+ & 0 \\ -a_- & 0 & 0 & a_-
\end{array}\right),
\label{dm}
\end{equation}
where we have used the basis $\{\ket{MM},\ket{M\bar M},\ket{\bar{M}M},\ket{\bar{M}\bar{M}}\}$
and $a_\pm = 1\pm e^{-2\lambda t}$. This expression is trace-preserving and is obtained by writing $\rho_{\mathcal{H}}(t)$ in the operator-sum representation and then performing the partial trace. We have neglected the effects of $CP$ violation; however, its inclusion would not affect our results significantly.

The approach used here can also be effectively applied to study observables of central importance in particle physics \cite{al}. 
For example, several important observables that are used to characterize meson decay processes can be developed using the above density matrix. Any physical observable of the neutral $B$-meson system is described by a suitable hermitian operator $\mathcal{O}$. Its evolution in time can be obtained by taking its trace with the density matrix $\rho(t)$,  ${\rm Tr}\,[ \mathcal{O}_f \, \rho(t)]$,  and from this standard results pertinent  to the meson systems can be derived. Hence, both quantum correlations as well as standard studies in particle physics can be carried out in a unified manner with the formalism used in this work.

\section{Interplay of quantum correlations in neutral meson systems}

Using the density matrix \eqref{dm}, we study the interplay of quantum correlations in our systems. 
We begin with Bell's inequality which was one of the first tools used to analyze and detect entanglement. Its physical content is that a system that can be described by a local realistic theory will satisfy this inequality. However, quantum mechanics seems to take delight in violating it \cite{aspect}! It is worth testing for such a violation in EPR-correlated $B$ and $K$ meson systems. Given a pair of qubits in the state $\rho$, we define the correlation matrix $T$ by $T_{mn}=Tr\left[\rho(\sigma_m\otimes \sigma_n)\right]$; let $u_i\ (i=1,2,3)$ be the eigenvalues of the matrix $T^{\dagger}T$. Then the Bell-CHSH inequality can be written $M(\rho)<1$, where $M(\rho)=\max(u_i+u_j)\ (i\neq j)$ \cite{hor}.  For the state \eqref{dm}, $M(\rho)$ is given by
\begin{equation}
 M(\rho)=(1+e^{-4 \lambda t}).
\end{equation}
Since Bell's inequality is not able to detect all possible entangled states, there is a need for some kind of measure which will quantify the amount of entanglement present in the system.  A well-known measure of entanglement is concurrence, which for a two-qubit system is equivalent to the entanglement of formation. For a mixed state $\rho$ of two qubits, the concurrence is \cite{wootters}
\begin{eqnarray}
 C=\max(\lambda_1-\lambda_2-\lambda_3-\lambda_4,0),
\end{eqnarray}
where $\lambda_i$ are the square root of the eigenvalues,  in decreasing order, 
of the matrix $\rho^{\frac{1}{2}}(\sigma_y\otimes \sigma_y)\rho^*(\sigma_y\otimes \sigma_y)\rho^{\frac{1}{2}}$ where $\rho$ is computed in the computational basis
$\{|00\rangle, |01\rangle, |10\rangle, |11\rangle\}$. 
For the state \eqref{dm}, concurrence has the simple analytical form
\begin{equation}
C=e^{-2\lambda t}.
\end{equation}
A similar result was also obtained in \cite{Bertlmann:2002wv}. The entanglement of formation can then be expressed as a monotonic function of concurrence C as
\begin{eqnarray}
 E_F=-\frac{1+\sqrt{1-C^2}}{2}\log_2(\frac{1+\sqrt{1-C^2}}{2})\nonumber\\-\frac{1-\sqrt{1-C^2}}{2}\log_2(\frac{1-\sqrt{1-C^2}}{2}).
\end{eqnarray}

As noted above, entanglement and quantum correlations need not be identical. Quantum discord attempts to reveal the quantum advantage over classical correlations. 
For the case of two qubits, 
geometric discord was shown \cite{dakic} to be
$D_{G}(\rho)=
\frac{1}{3}[\|\vec{x}\|^{2}+\|T\|^{2}-\lambda_{max}(\vec{x}\vec{x}^{\dagger}+ T T^{\dagger})]$ where $T$ is the correlation matrix defined above, $\vec x$ is the vector whose components are $x_{m}=\mathrm{Tr}(\rho(\sigma_{m}\otimes {{\mathbb I}}_{2}))$, and $\lambda_{max}(M)$ is the maximum eigenvalue of the matrix $M$. Here it can be seen that $D_{G}(\rho)=M(\rho)/3$, in consistence with the findings in \cite{ab12}. 

Teleportation provides an operational meaning to entanglement. The classical fidelity of teleportation in the absence of entanglement is $2/3$. Thus, whenever $F_{\max}>2/3$,  teleportation is possible. Interestingly, this does not rule out the possibility of entangled states that, while they do not violate Bell's inequality, can nonetheless be useful for teleportation.
$F_{\max}$ is easily computed in terms of the eigenvalues $\{u_i\}$ of $T^\dagger T$ mentioned earlier: it is
$F_{\max}=\frac{1}{2}\left(1+\frac{1}{3}N(\rho)\right)$ where $N(\rho)=\sqrt{u_1}+\sqrt{u_2}+\sqrt{u_3}$. The calculation of $F_{\max}$ for the state \eqref{dm} gives
\begin{equation}
F_{\max} = \frac{1}{12}\Big[6+2e^{-2\lambda t} +\sqrt{2}\sqrt{\alpha-\sqrt{\beta}} + \sqrt{2}\sqrt{\alpha+\sqrt{\beta}}\Big],
\end{equation}
where
\begin{eqnarray}
\alpha &=& 1 + \cosh(4 \lambda t)-\sinh(4 \lambda t),\\
\beta &=& 3 - 2\alpha + \cosh(8 \lambda t)-\sinh(8 \lambda t).
\end{eqnarray}
Thus we see that for $\lambda=0$, $E_F=F_{max}$.
A useful inequality involving $M(\rho)$ and $F_{\max}$ is \cite{hor}:
\begin{equation}
F_{max}\ge \frac12\left(1+\frac13 M(\rho)\right)\ge\frac23\ {\rm if}\ M(\rho)>1.
\label{inequality}
\end{equation}
We will see that this can be violated for unstable systems.

To take into account the effect of decay in the systems under study, the various correlations are modified by the probability of survival of the pair of particles up to that time, which can be shown to be $e^{-2\Gamma t}$, where $\Gamma$ is the meson decay width. 
For the $K$ meson, $\Gamma=\frac{1}{2}(\Gamma_S + \Gamma_L)$ (where $\Gamma_S$ and $\Gamma_L$ are the decay widths of short and long neutral kaon states, respectively); its value is $5.59 \times 10^{9}\,\rm s^{-1}$ \cite{pdg}. The decay widths for $B_d$ and $B_s$ mesons are $6.58 \times 10^{11}\, \rm s^{-1}$ and $6.61 \times 10^{11}\, \rm s^{-1}$, respectively \cite{hfag}. 

\begin{figure*}[htp] 
\centering
\begin{tabular}{ccc}
\includegraphics[width=55mm]{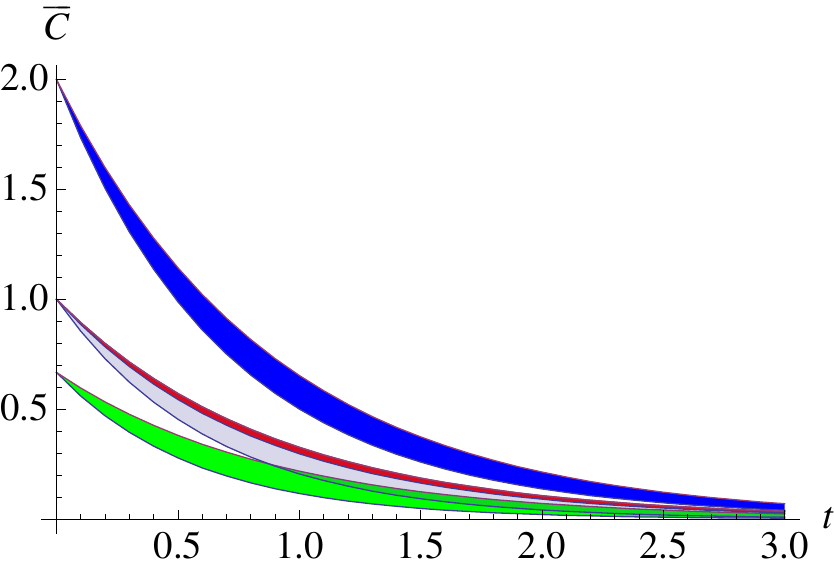}&
\includegraphics[width=55mm]{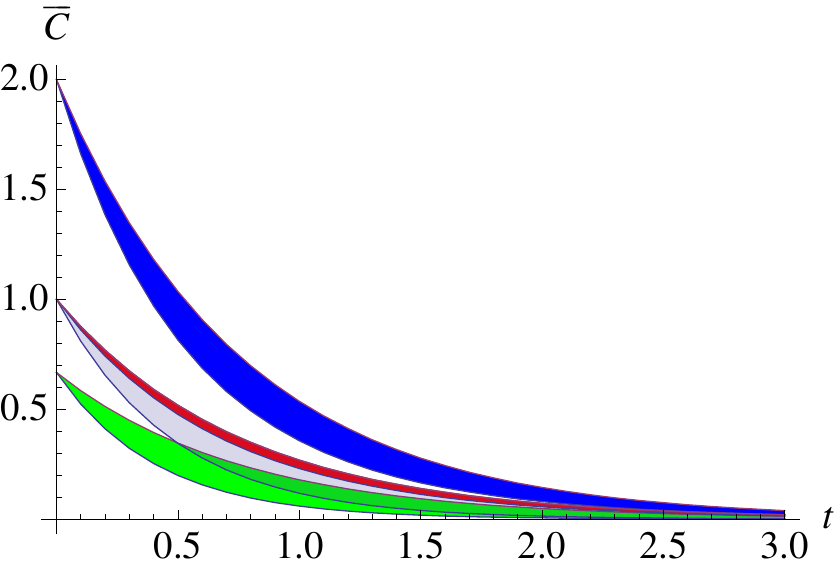}&
\includegraphics[width=55mm]{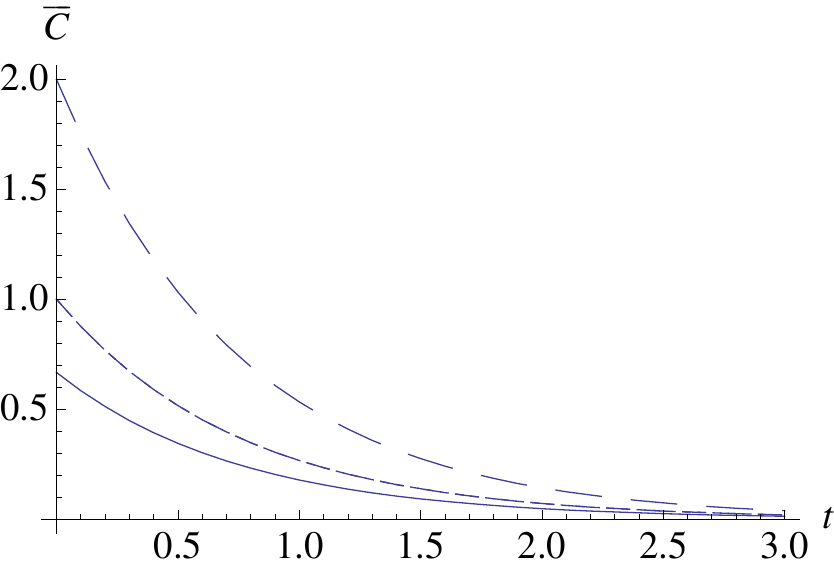}\\
\end{tabular}
\caption{(color online) Average correlation measures as a function of time $t$.
 The left, middle and right panels correspond to the correlations of a $K\bar{K}$, $B_d\bar{B_d}$ and $B_s\bar{B_s}$ pair created at $t=0$, respectively. The four correlation measures are (top to bottom): $M(\rho)$ (Bell's inequality; blue band), $F_{\max}$ (teleportation fidelity; red band), $E_F$ (entanglement of formation; grey band) and $D_G$ (geometric discord; green band). For $K\bar{K}$ pairs, left panel, time is in units of $10^{-10}$ seconds whereas for the $B_d\bar{B_d}$ and $B_s\bar{B_s}$ pairs, time is in units of $10^{-12}$ seconds (in all cases, the approximate lifetime of the particles). In the left and middle panels, the bands represent the effect of decoherence corresponding to a $3\sigma$ upper bound on the decoherence parameter $\lambda$. The right panel has no such bands because there is currently no experimental evidence for decoherence in the case of $B_s$ mesons; for this case, $F_{\max}=E_F$.
}
\label{kbmeson}
\end{figure*}

The parameter $\lambda$ models the effect of decoherence. In the case of the $K$ meson system, its value has been obtained by the KLOE collaboration by studying the interference between the initially entangled kaons and the decay product in the channel $\phi \to K_S K_L \to \pi^+ \pi^- \pi^+ \pi^-$ \cite{Ambrosino:2006vr}. The value of $\lambda$ is restricted to $1.58 \times 10^9\, {\rm s^{-1}}$ at $\rm 3 \sigma$. In the case of $B$ meson systems, the decoherence parameter is determined by using time-integrated dilepton events. The value of $\lambda$, for $B_d$ mesons, is determined by the measurement of ratio of the total same-sign to opposites-sign semileptonic rates, $R_d$, and is 
restricted to $2.82 \times 10^{11}\, {\rm s^{-1}}$ at $\rm 3 \sigma$  \cite{Bertlmann:2001iw}. However there has been no experimental update for $R_d$ since \cite{R-chi} two decades. For $B_s$ mesons, to the best of our knowledge, there is no experimental information about $\lambda$ so we will take it to be zero in what follows. The results for $K$ and $B$ meson systems are shown in Fig.~\ref{kbmeson}. 

The nonclassicality of quantum correlations, in the neutral mesons, can be characterized in terms of nonlocality (which is the strongest condition), entanglement, teleportation fidelity or weaker nonclassicality measures like quantum discord. The fall in the pattern of the average value of these correlations, as shown in  Fig.~\ref{kbmeson},  are in accord with the fact that here we are dealing with unstable particles, which decay with time. From the left panel of Fig.~\ref{kbmeson}, one can see that until about 50\% of the average life time of $K_S$ meson in the presence of decoherence and about 60\% in its absence, Bell's inequality is violated. This means that, in the conventional sense, until this time, the time evolution cannot be simulated by any local realistic theory. However, we find that even if Bell's inequality is violated ($M(\rho)>1$), the teleportation fidelity $F_{\max}$ could be below the classical value of 2/3. 
For example, from the left panel of Fig.~\ref{kbmeson}, it is seen that, in the absence of decoherence, $F_{\max}$ drops below 2/3 as $M(\rho)$ drops below 1.3, in violation of the inequality \eqref{inequality} \cite{hor} according to which the cutoff is $M(\rho)=1$. This violation is slightly reduced, but nonetheless still occurs, even in the presence of decoherence, starting at $M(\rho)\simeq1.2$. This is consistent with the degradation of correlations with decoherence.

The violation can be understood mathematically since the individual measures of correlation are modulated by a factor $e^{-2\Gamma t}$. Since \eqref{inequality} contains constant terms which are not modulated, it is affected by these modulations.

Thus we see that the study of quantum correlations in unstable systems is nontrivially different from their stable counterparts. However, there are some similarities as well. In particular, the average geometric discord is always bounded from above by the average entanglement of formation. This is consistent with the fact that discord is a weaker measure of quantum correlations compared to entanglement \cite{ab12}. From the middle and right panel of Fig.~\ref{kbmeson}, we see that the above conclusions hold also for $B_d$ and $B_s$ meson systems, respectively. 
Here we would like to point out that the nontrivial differences between the meson systems and their stable counterparts is only due to the decaying nature of the system and not due to oscillations. This is borne out by the fact that a study of quantum correlations in a stable, oscillating subatomic system such as a neutrino, shows no such deviation \cite{asu}.

From Fig.~\ref{kbmeson}, we see that the spread in the various correlations, corresponding to $\rm 3\sigma$ upper bound on the decoherence parameter $\lambda$, is more prominent for the $B_d$, compared to the $K$ meson system. This is because of the choice of time scales used in the plots. In the case of $K$ mesons, the time scale is $10^{-10}\rm s$, which is roughly the average lifetime of the $K_S$ mesons; whereas in the case of $B_d$ and $B_s$ mesons, the time scale is picosecond, which is roughly the average life time of these mesons.  As can be seen from \eqref{dm}, the coherences in the system, of which the correlations would be a function, depend upon the evolution time $t$ and the decoherence parameter $\lambda$, as a function of $\lambda t$. Thus from the values of $\lambda$'s for these systems (see below \eqref{dm}), it can be easily inferred that the effect of decoherence is more prominent, in the chosen time scales, for the $B_d$ meson as compared to the $K$ meson system. This is consistent with Fig.~\ref{kbmeson}.

\section{Conclusions}

To conclude, in this work we have studied  a number of aspects of quantum correlations in correlated neutral meson systems, {\em viz.} neutral $K$, $B_d$ and $B_s$ mesons. 
This is nontrivial given the fact that the mesons decay with time. This was accomplished by using the semigroup formalism, well known in the study of open quantum systems.
We have also studied the impact of decoherence on various correlation measures in these systems. We found that the quantum correlations here can be nontrivially different from their stable counterparts. This was made explicit by the interplay between Bell's inequality and teleportation fidelity. On average, Bell's inequality in these correlated-meson systems is violated for about half of the meson lifetime. One particularly surprising result is that teleportation fidelity does not exceed the classical threshold of 2/3 for all Bell's inequality violations. 
This behavior, not seen in stable systems, is interesting since one of the cornerstones in the field of quantum information is the interplay between Bell's inequality and teleportation fidelity. 
This surprising behavior is due to the decaying nature of the parent system and not due to flavor oscillations. There are some similarities as well. This is particularly seen by the fact that entanglement provides an upper bound to discord at all times.  This is the crux that provides insight into the differences/similarities in quantum correlations in these meson systems in comparison to their stable counterparts. Thus this work provides important insight into foundational issues in the context of quantum mechanics of unstable subatomic particles, and presumably other unstable systems as well.

On the experimental front, our results can impact probing the nonlocality of $B$ mesons, in particular the Bell's inequality, which, at present, leads to difficulties due to the lack of active measurements. Also, the measurement of decoherence parameter in the $B_s$ mesons can provide insight into the fundamental nature of these systems. This work could hopefully motivate future experiments (or reanalysis of past experimental results such as the wealth of results from decommissioned $B$ factories) to probe these issues.

\section*{Acknowledgments}
 
SB and AKA thank R. Srikanth, A. Pathak and S. Uma Sankar for useful comments on the manuscript. The work of RM is supported by the Natural Science and Engineering Research Council of Canada and by the Fonds de recherche du Qu\'ebec -- Nature et technologies.


\begin{thebibliography}{10}

\bibitem{Bramon:1998nz} 
  A.~Bramon and M.~Nowakowski,
  Phys.\ Rev.\ Lett.\  {\bf 83}, 1 (1999)
  [hep-ph/9811406].
  
  \bibitem{Bramon:2001tb} 
  A.~Bramon and G.~Garbarino,
  Phys.\ Rev.\ Lett.\  {\bf 88}, 040403 (2002)
  [quant-ph/0108047].
  
  \bibitem{Bramon:2002yg} 
  A.~Bramon and G.~Garbarino,
  Phys.\ Rev.\ Lett.\  {\bf 89}, 160401 (2002)
  [quant-ph/0205112].

\bibitem{Bertlmann:2002wv} 
  R.~A.~Bertlmann, K.~Durstberger and B.~C.~Hiesmayr,
  Phys.\ Rev.\ A {\bf 68}, 012111 (2003)
  [quant-ph/0209017].

  \bibitem{Bramon:2004zp} 
  A.~Bramon, G.~Garbarino and B.~C.~Hiesmayr,
  Phys.\ Rev.\ A {\bf 69}, 062111 (2004)
  [quant-ph/0402212].

\bibitem{Hiesmayr:2007he} 
  B.~C.~Hiesmayr,
  Eur.\ Phys.\ J.\ C {\bf 50}, 73 (2007).

  \bibitem{Bertlmann:2004cr} 
  R.~A.~Bertlmann, A.~Bramon, G.~Garbarino and B.~C.~Hiesmayr,
  Phys.\ Lett.\ A {\bf 332}, 355 (2004)
  [quant-ph/0409051].
  
  \bibitem{active}
  A.~Go {\it et al.}  [Belle Collaboration],
  Phys.\ Rev.\ Lett.\  {\bf 99}, 131802 (2007)
  [quant-ph/0702267 [QUANT-PH]].
  
  \bibitem{Blasone:2007vw} 
  M.~Blasone, F.~Dell'Anno, S.~De Siena and F.~Illuminati,
  Europhys.\ Lett.\  {\bf 85}, 0002 (2009)
  [arXiv:0707.4476 [hep-ph]].

\bibitem{Blasone:2007wp} 
  M.~Blasone, F.~Dell'Anno, S.~De Siena, M.~Di Mauro and F.~Illuminati,
  Phys.\ Rev.\ D {\bf 77}, 096002 (2008)
  [arXiv:0711.2268 [quant-ph]].
  
  \bibitem{AmelinoCamelia:2010me} 
  G.~Amelino-Camelia, F.~Archilli, D.~Babusci, D.~Badoni, G.~Bencivenni, J.~Bernabeu, R.~A.~Bertlmann and D.~R.~Boito {\it et al.},
  Eur.\ Phys.\ J.\ C {\bf 68}, 619 (2010)
  [arXiv:1003.3868 [hep-ex]].
 
  
  \bibitem{Hiesmayr:2011na} 
  B.~C.~Hiesmayr, A.~Di Domenico, C.~Curceanu, A.~Gabriel, M.~Huber, J.~A.~Larsson and P.~Moskal,
  Eur.\ Phys.\ J.\ C {\bf 72}, 1856 (2012)
  [arXiv:1111.4797 [quant-ph]].
  
  
\bibitem{Blasone:2014jea} 
  M.~Blasone, F.~Dell'Anno, S.~De Siena and F.~Illuminati,
  Europhys.\ Lett.\  {\bf 106}, 30002 (2014)
  [arXiv:1401.7793 [quant-ph]].
  
  \bibitem{asu}
  A.~K.~Alok, S.~Banerjee and S.~U.~Sankar,
  arXiv:1411.5536 [quant-ph].

  
\bibitem{louisell}
 W. H. Louisell, {\it Quantum Statistical Properties of Radiation}
(Wiley, New York, 1973). 

\bibitem{bp}
H-P. Breuer and F. Petruccione, {\it The Theory of Open Quantum Systems} (Oxford University Press, 2003).

\bibitem{nc} M. A. Nielsen and I. L. Chuang, {\it Quantum Computation and Quantum Information} (Cambridge University Press, 2010).
 
  
\bibitem{path}
U. Weiss, {\it Quantum Dissipative Systems} (World Scientific,
Singapore, 2001).

\bibitem{hor}  R.  Horodecki,  P.  Horodecki and M. Horodecki,  Phys. Lett. A  {\bf 222}, 21 (1996).

\bibitem{sb1}
I. Chakrabarty, S. Banerjee and N. Siddharth, Quantum Information and Computation, {\bf 11}, 0541 (2011). 


\bibitem{Weisskopf:1930au} 
  V.~Weisskopf and E.~P.~Wigner,
  Z.\ Phys.\  {\bf 63}, 54 (1930).

\bibitem{Weisskopf:1930ps} 
  V.~Weisskopf and E.~Wigner,
  Z.\ Phys.\  {\bf 65}, 18 (1930).
  
  \bibitem{peierls-1}
R. E. Peierls, {\it Proceedings of the Glasgow Conference}, Glasgow, Scotland,  296-299, (1954).
  
\bibitem{peierls-2}
R. E. Peierls, Proc. R. Soc. London {\bf 253A}, 16 (1960).
  
\bibitem{khalfin}
L. Khalfin, Sov. Phys. JETP {\bf 6}, 1053 (1958); JETP Lett. {\bf 8}, 65 (1968).
  
\bibitem{prig}
T. Petrosky and I. Prigogine, Proc. Nat. Acad. Sci. U. S. A. {\bf 93}, 9393 (1993); I. Antoniou and I. Prigogine, Physica A {\bf 173}, 175 (1991).
  
\bibitem{kabir} 
  P.~K.~Kabir and A.~Pilaftsis,
  Phys.\ Rev.\ A {\bf 53}, 66 (1996)
  [hep-ph/9509349].
  
\bibitem{sud}
E.~C.~G.~ Sudarshan, Charles B. Chiu and G. Bhamathi, {\it Advances in Chemical Physics XCIX}, John Wiley \& Sons, Inc., 121-210, (1997).

\bibitem{Ellis:1992dz} 
  J.~R.~Ellis, N.~E.~Mavromatos and D.~V.~Nanopoulos,
  Phys.\ Lett.\ B {\bf 293}, 142 (1992)
  [hep-ph/9207268].


\bibitem{Huet:1994kr} 
  P.~Huet and M.~E.~Peskin,
  Nucl.\ Phys.\ B {\bf 434}, 3 (1995)
  [hep-ph/9403257].
  
\bibitem{benatti}
  F.~Benatti and R.~Floreanini,
  Nucl.\ Phys.\ B {\bf 511}, 550 (1998)
  [hep-ph/9711240].

\bibitem{benatti2} 
  F.~Benatti, R.~Floreanini and R.~Romano,
  Nucl.\ Phys.\ B {\bf 602}, 541 (2001)
  [hep-ph/0103239].
  
  \bibitem{Urbanowski:2014tra} 
  K.~Urbanowski,
  Open Systems \& Information Dynamics {\bf 20}, 1340008 (2013)
  [arXiv:1408.2211 [quant-ph]].
  
  \bibitem{rajgopal}
A.~Weron, A.~K.~Rajagopal and K.~Weron, Phys. Rev. A {\bf 31}, 1736 (1985).
  
  \bibitem{nagy}
B. Sz.-Nagy and C. Foias, Acta Sci. Math. 21, 251 (1960); B. Sz.-Nagy and C. Foias, {\it Harmonic Analysis of Operators on Hilbert Spaces},  North-Holland, Amsterdam, (1970).

\bibitem{caban} 
  P.~Caban, J.~Rembielinski, K.~A.~Smolinski and Z.~Walczak,
  Phys.\ Rev.\ A {\bf 72}, 032106 (2005)
  [quant-ph/0506183]; 
  P.~Caban, J.~Rembielinski, K.~A.~Smolinski, Z.~Walczak and M.~Wlodarczyk,
  Phys.\ Lett.\ A {\bf 357}, 6 (2006)
  [quant-ph/0603169];
  P.~Caban, J.~Rembielinski, K.~A.~Smolinski and Z.~Walczak,
  Phys.\ Lett.\ A {\bf 363}, 389 (2007).
  

\bibitem{alicki} R. Alicki and K. Lendi, \textit{Quantum Dynamical Semigroups and Applications} (Lect. Notes
Phys. 717 (Springer, Berlin Heidelberg 2007)).


\bibitem{Hawking:1982dj} 
  S.~W.~Hawking,
  Commun.\ Math.\ Phys.\  {\bf 87}, 395 (1982).
  
  
  \bibitem{epr} A. Einstein, {\em et al.}, Phys. Rev.  \textbf{47}, 777 (1935).

\bibitem{schr} E. Schr{\"o}dinger, "Die gegenwŠrtige Situation in der Quantenmechanik", Naturwissenschaften \textbf{23}, 807 (1935); ibid., 823 (1935); ibid., 844 (1935); Proc. Cambridge Phil. Soc. \textbf{31}, 555 (1935); ibid., \textbf{32}, 446 (1936).

\bibitem{bell} J. S. Bell, Physics \textbf{1}, 195 (1964).

\bibitem{chsh} J. F. Clauser, and A. Shimony, Rep. Progr. Phys. \textbf{41}, 1881 (1978).


\bibitem{ollivier} H. Ollivier, and W. H. Zurek, Phys. Rev. Lett. \textbf{88}, 017901 (2001).

\bibitem{henderson} L. Henderson, and V. Vedral, J. Phys. A \textbf{34}, 6899 (2001).

\bibitem{luo} S. Luo, Phys. Rev. A \textbf{77}, 042303 (2008).

\bibitem{girolami1} D. Girolami, and G. Adesso, Phys. Rev. A \textbf{83}, 052108 (2011).

\bibitem{huang} Y. Huang, New J. Phys. {\bf 16}, 033027 (2014). 

\bibitem{dakic} B. Dakic, V. Vedral and C. Brukner, Phys. Rev. Lett. \textbf{105}, 190502 (2010).

\bibitem{ben}  C.  H.  Bennett,  G.  Brassard,  C.  Crepeau,  R.  Jozsa,  A.  Peres and W.  K. Wootters,  Phys. Rev. Lett.
{\bf 70}, 1895 (1993).

\bibitem{al}
R. Alicki, Rep. on Math. Phys. {\bf 14}, 27 (1978).
  

\bibitem{aspect}
A. Aspect, J. Dalibard and G. Roger, Phys. Rev. Lett. {\bf 49}, 1804 (1982);
A. Aspect, P. Grangier and G. Roger, Phys. Rev. Lett. {\bf 49}, 91 (1982);
A. Aspect, P. Grangier and G. Roger, Phys. Rev. Lett. {\bf 47}, 460 (1981).

\bibitem{wootters} S. Hill, and W. K. Wootters, Phys. Rev. Lett. \textbf{78}, 5022 (1997);
W. K. Wootters, Phys. Rev. Lett. \textbf{80}, 2245 (1998).

\bibitem{ab12} S. Adhikari and S. Banerjee, Phys. Rev. A {\bf 86}, 062313 (2012).

  \bibitem{pdg} 
  J.~Beringer {\it et al.}  [Particle Data Group Collaboration],
  Phys.\ Rev.\ D {\bf 86}, 010001 (2012).

\bibitem{hfag}
  Y.~Amhis {\it et al.}  [Heavy Flavor Averaging Group Collaboration],
  arXiv:1207.1158 [hep-ex].
  
\bibitem{Ambrosino:2006vr} 
  F.~Ambrosino {\it et al.}  [KLOE Collaboration],
  Phys.\ Lett.\ B {\bf 642}, 315 (2006)
  [hep-ex/0607027].

\bibitem{Bertlmann:2001iw} 
  R.~A.~Bertlmann and W.~Grimus,
  Phys.\ Rev.\ D {\bf 64}, 056004 (2001)
  [hep-ph/0101160].
  
    \bibitem{R-chi} 
  H.~Albrecht {\it et al.}  [ARGUS Collaboration],
  Phys.\ Lett.\ B {\bf 324}, 249 (1994);
   J.~E.~Bartelt {\it et al.}  [CLEO Collaboration],
  Phys.\ Rev.\ Lett.\  {\bf 71}, 1680 (1993).
  


\end{thebibliography}
\end{document}